\documentclass[12pt]{article}
\usepackage{graphicx,amsfonts,amsbsy}

\newcommand*\scom[2]{[  #1 \, \stackrel{\star}{,} \, #2  ]}

\def\id{\mathrm{id}}

\def\ed{\mathrm{d}}

\def\BC{{\mathbb{C}}}
\def\BR{{\mathbb{R}}}
\def\BZ{{\mathbb{Z}}}

\newcommand*{\Ax}{\mathfrak{A}}

\newcommand*{\hdelta}{\hat\delta}
\def\AA{{\cal A}}

\def\DD{{\cal D}}

\def\EE{{\cal E}}

\def\eq#1\en{\begin{equation}#1\end{equation}}
\def\eqa#1\ena{\begin{eqnarray}#1\end{eqnarray}}
\newcommand*\nn{\nonumber \\}

\newlength{\vscaling} \newlength{\hscaling}
  \newcommand*\tib[2]{{\em #1}, #2}

\begin{document}
\begin{titlepage}
\rightline{LMU-TPW 01-06}
\vfill

{\LARGE\bf
\begin{center}
Noncommutative line bundle and Morita equivalence
\end{center}
}

\begin{center}

\vfill

{{\bf Branislav Jur\v co, Peter Schupp and Julius Wess
}}

 \vskip 0.5 cm

Universit\"at M\"unchen,
Sektion Physik\\ Theresienstr.\ 37,
80333 M\"unchen, Germany
\end{center}
\vfill
\begin{abstract} 
Global properties of abelian noncommutative gauge theories based on
$\star$-products which
are deformation quantizations of arbitrary Poisson structures
are studied.
The consistency condition for finite noncommutative gauge transformations
and its explicit solution in the abelian case are given.
It is shown that the local existence of invertible covariantizing
maps (which are closely related
to the Seiberg-Witten map) leads naturally to the notion of a noncommutative
line bundle with noncommutative
transition functions. We introduce the space of sections of such a line
bundle and explicitly show that it is a projective module.
The local covariantizing maps define a new star
product $\star'$ which is shown to be
Morita equivalent to $\star$. 
\end{abstract}
\vfill
\hrule
\vskip 5pt
\noindent
{\footnotesize\it e-mail:
\parbox[t]{.8\textwidth}{jurco\,,\,schupp\,,%
\,wess@theorie.physik.uni-muenchen.de}}
\end{titlepage}\vskip.2cm

\newpage

\setcounter{page}{1}

\section{Introduction}

Noncommutativity plays a prominent role in physics ever since the birth 
of quantum mechanics.
When trying to replace the notions and concepts of 
commutative geometry in the more general noncommutative framework the basic 
strategy is to not consider the manifold itself but rather the algebra of 
functions on it.
In the noncommutative realm the algebra of functions is replaced
by an arbitrary associative algebra $\Ax$. In the spirit of
the Gel'fand-Naimark theorem
we shall keep referring to the elements of $\Ax$ as ``functions
on the noncommutative space.''
Similarly, the celebrated Serre-Swan theorem allows us to replace
the notion of
vector bundle by the one of projective module of sections.
Noncommutative Yang-Mills theories are naturally formulated in terms of 
projective modules \cite{Connes}.  
In string theory $D$-branes are a realization of
vector bundles on quite general sub-manifolds of space-time.
In the presence of background fields these become noncommutative,
see \cite{CDS,SW} and references therein.

In this paper we deal with the situation when our noncommutative space $\Ax$ is a 
deformation quantization of some Poisson manifold $M$. Since every deformation 
quantization originates from one of Kontsevich type, we can restrict ourselves 
to this case. Based on our previous studies of (abelian) noncommutative gauge 
theories \cite{JS,MSSW,JSW1,JSW2} we give the explicit form of finite
noncommutative gauge transformations and develop the concept of a
noncommutative line bundle, in the sense of 
deformation quantization, that is closer to the spirit of algebraic topology then 
a formulation
in terms of projective modules.

The type of noncommutative gauge theory that we take as a
starting point in this article
can be based on a few basic ideas
which we shall briefly review below: the concept of
covariant functions, the requirement of locality (in the gauge potential)
 and accompanying
gauge equivalence and consistency conditions.

It is natural to introduce local 
gauge transformations of a field $\Psi$ 
on a noncommutative Space $\Ax$ in analogy to the 
commutative case by\footnote{Here and in the following we use capital letters to denote
noncommutative quantities.}
\eq
\hdelta \Psi = i \Lambda \star \Psi, \qquad \Psi,\Lambda \in \Ax.
\label{deltapsi}
\en
Multiplying a field from the right by a function yields a new
field that transforms again according to (\ref{deltapsi}).
This is not the case, however,
if we multiply a field from the \emph{left}
by a function $f$, simply because the local gauge parameter $\Lambda$ will not
in general commute with $f$:
\[
\hdelta (f\star\Psi) = i f\star\Lambda \star \Psi
\neq i \Lambda \star (f\star\Psi);
\qquad (\hdelta f = 0).
\]
Consider the case of a coordinate function $x^i$ (i.e.\ a generator
of $\Ax$):
In complete analogy to ordinary gauge theory (where the gauge
parameter does not commute with derivatives) one needs to introduce
noncommutative gauge potentials $A^i$ and covariant coordinates $X^i = x^i + A^i$
with transformation properties~\cite{MSSW}
\eq
\hdelta X^i = i\scom{\Lambda}{X^i}\quad\Leftrightarrow
\quad
\hdelta A^i = i\scom{\Lambda}{x^i}
+ i\scom{\Lambda}{A^i}.
\en
Covariant functions in general are introduced  via an invertible
gauge-de\-pen\-dent map
$\DD:\Ax\rightarrow\Ax$
that transforms under gauge transformations
such that $\hdelta (\DD f)  = i\scom{\Lambda}{\DD f}$. The 
gauge-de\-pen\-dent map $\AA = \DD - \id$ plays the role of a generalized
noncommutative gauge potential. The more familiar
noncommutative gauge potentials $A^i$ are obtained by evaluating $\AA$ on
a coordinate function $x^i$.
The covariant coordinates $X^i$ generate
a new algebra $\Ax'$ with product~$\star'$ that can be computed
on any pair of functions via $\DD(f \star' g) = \DD f \star \DD g$. In this way
we have a (physically
desirable) back-reaction of the gauge fields on the noncommutative
space.

If a classical (commutative) limit of the theory exists, as is for instance
the case, when the noncommutative structure is given by a star
product,  one may ask about the relation between the noncommutative
and corresponding classical gauge fields.
It turns out that one can find maps $A[a]$, $\Lambda_\lambda[a]$
that express the noncommutative
gauge field and gauge parameter in terms of their classical counterparts
$a$, $\lambda$
such that a \emph{gauge equivalence condition}
\eq
A[a+ \delta_\lambda a] = A[a] + \hdelta_{\Lambda_\lambda[a]} A[a] + o(\lambda^2)
\en
holds~\cite{SW}: Classical gauge transformations
$\delta_\lambda$ induce noncommutative
ones.
These and similar mappings, require that \emph{all noncommutative
quantities  are local functions of the classical gauge potential and
its derivatives}.
Extending the gauge equivalence condition to fields $\Psi$ that transform
in the fundamental
\eq
\Psi[\psi+ \delta_\lambda\psi,a + \delta_\lambda a] = \Psi[\psi,a] + i \Lambda_\lambda[a]
\star \Psi[\psi,a] + o(\lambda^2)
\en
on can derive a \emph{consistency condition} that involves only the gauge
parameters:
\eq
\scom{\Lambda_\alpha[a]}{\Lambda_\beta[a]} + i\delta_\alpha \Lambda_\beta[a]
- i\delta_\beta \Lambda_\alpha[a] =  \Lambda_{[\alpha,\beta]}[a].
\en
This condition is of central importance in the present
work, since it defines a noncommutative group law as we shall see.
(We have written the nonabelian version of the consistency condition --
in the abelian case $[\alpha,\beta] = 0$.)

We are particularly interested in noncommutative structures that
are given by a star product, i.e.\ an associative algebra
$\Ax = (C^\infty(M)[[\hbar]],\star)$
that is the deformation quantization of a Poisson structure
over some mani\-fold $M$.
For an arbitrary Poisson structure $\theta$
a star product $\star$ exists and
can be expressed in terms of the Kontsevich formality map
\cite{Kontsevich}.
In~\cite{JS,JSW1,JSW2} we have used this map to construct explicit solutions
$\DD_{[a]}$ and $\Lambda_\lambda[a]$ 
to
the gauge equivalence
\eq
\DD_{[a + \delta_\lambda a]}(f) = \DD_{[a]}(f) + i
\scom{\Lambda_\lambda[a]}{\DD_{[a]}(f)} \label{gequi}
\en
and consistency condition
\eq
\scom{\Lambda_\alpha[a]}{\Lambda_\beta[a]} + i\delta_\alpha \Lambda_\beta[a]
- i\delta_\beta \Lambda_\alpha[a] =  0. \label{consist}
\en
(For the following the existence of 
$\star$, $\DD_{[a]}$ and $\Lambda_\lambda[a]$ that satisfy
(\ref{gequi}) and (\ref{consist}) 
is more important than the explicit form of these objects.)
The map $\DD_{[a]}$ is formally invertible and defines an equivalent star
product~$\star'$ via $\DD_{[a]}(f\star' g) =  \DD_{[a]} f \star  \DD_{[a]} g$
on the patch where the local gauge potential $a$ is defined.
The new star product $\star'$ itself can be defined globally
since it only depends on $a$ via its gauge-invariant
field strength $f = da$: The star product $\star'$ is the
deformation quantization by Kontsevich's formula of the Poisson structure
$\theta'$ which, for some choice of local coordinates and using matrix notation
has the explicit form $\theta' = \sum_{n=0}^\infty (-1)^n \theta (f
\theta)^n$.\footnote{In general, the local gauge potential
$a$ can be a formal power series
in the deformation parameter~$\hbar$. To conform with the mathematics literature
it can be taken to start with a term of order~$\hbar$; the expression for $\theta'$ is
then also a formal power series.}
The star products
$\star$ and $\star'$ are ``patch-wise'' equivalent. We will show in this
article that the corresponding algebras $\Ax$, $\Ax'$ are in fact Morita
equivalent.\footnote{See~\cite{Schwarz} for the relevance of Morita equivalence
in string theory.}

The Kontsevich formality maps can be computed explicitly on open subsets of $\BR^n$
with diagrammatic techniques that resemble those of Feynman diagrams;
they can then be consistently globalized using maps
that in fact closely resemble the  $\DD_{[a]}$. In the following we shall
assume that this has already be done, i.e., we have a globally defined
star product $\star$.
In our previous work we have claimed that our formulas for $\DD_{[a]}$,
$\Lambda_\lambda[a]$ and also
the noncommutative gauge potential can be globalized by
noncommutative gauge transformations between patches; this will be
made precise in this article.

\section{Finite gauge transformations}
\label{sec2}

Let us consider finite classical gauge transformations
\eq
\psi \mapsto \psi_g = g \psi, \qquad a \mapsto a_g =  a+ i g \ed g^{-1} .
\en
(In the nonabelian case the latter would read $a \mapsto a_g =
g a g^{-1} + i g \ed g^{-1}$.)
Corresponding finite versions of the
gauge equivalence conditions for the covariant functions
and fields are
\eq
\DD_{[a_g]}(f) \star G_g[a] = G_g[a] \star \DD_{[a]}(f),
\label{equivalence}
\en
\eq
\Psi[\psi_g,a_g] = G_g[a] \star \Psi[\psi,a],
\en
where $\DD_{[a]}$ is an invertible map. 
\emph{Any} covariant function, i.e.\ a $f_{[a]}$ which under a gauge transformation
$a \mapsto a_g$ transforms as
\eq
f_{[a_g]} = G_g[a] \star f_{[a]} \star (G_g[a])^{-1},
\en
is necessarily
of the form $\DD_{[a]}(f')$ with an invariant $f'$.
In fact, using equation (\ref{equivalence})  in the form
\eq
\mathrm{Ad}_\star G_g[a] = \DD_{[a_g]} \circ \DD^{-1}_{[a]}
\en
it is easy to see that
$f' \equiv \DD_{[a]}^{-1}(f_{[a]})$ is invariant.

Evaluating two consecutive gauge transformations $g_1$, $g_2$ on
$\Psi = \Psi[\psi,a]$, 
$\Psi \stackrel{g_1}{\mapsto} G_{g_1}[a] \star \Psi
\stackrel{g_2}{\mapsto} G_{g_1}[a_{g_2}] \star G_{g_2}[a]\star \Psi
= G_{g_1 \cdot g_2}[a]\star \Psi$,  we obtain the
gauge consistency condition
\eq
G_{g_1}[a_{g_2}] \star G_{g_2}[a] = 
G_{g_1\cdot g_2}[a]. \label{consistency}
\en
One should note the appearance of the shifted gauge
potential $a_{g_2}$ in this formula -- at first sight this seems
to preclude the use of (\ref{consistency}) to define
transition functions of a noncommutative bundle, but
it turns out that this is exactly what is needed.
The finite noncommutative gauge transformation corresponding to
$g = e^{i\lambda}$ can be computed by evaluating $e^{\delta_\lambda}$
on a field $\Psi$ using the gauge equivalence condition
\eq
\delta_\lambda \Psi[\psi,a] = i \Lambda_\lambda[a] \star \Psi[\psi,a]
\en
repeatedly. One has to be careful though, because $\delta_\lambda$
not only affects $\Psi$ but also $a$ in $\Lambda_\lambda[a]$.
This difficulty can be bypassed without having to resort to
path-ordered exponentials with the following trick:
\eq
(-\delta_\lambda + i \Lambda_\lambda[a]\star)\Psi = 0
\quad\Rightarrow\quad
e^{\delta_\lambda} \Psi[\psi,a] =
\left(e^{\delta_\lambda} e^{-\delta_\lambda + i \Lambda_\lambda[a]\star}
\right) \Psi[\psi,a].
\en
Using the Baker-Campbell-Hausdorff formula the exponentials
can be combined and one finds that there
are no free $\delta_\lambda$'s acting  on the $\Psi[\psi,a]$. Thus
$(G_{(e^{i\lambda})}[a]) \star
= \left(e^{\delta_\lambda} e^{-\delta_\lambda + i \Lambda_\lambda[a]\star}
\right)$.
Using that the gauge transformation $\delta_\lambda$ in this formula
is actually just a shorthand for the functional expression
\eq
\delta_\lambda = \int \delta_\lambda(a(x)) \frac{\delta}{\delta a(x)} dx
\en
and noting that for suitably chosen star product we can always
insert a trailing $\star$ in this integral,\footnote{For constant $\theta$ this
is clear by partial integration for the general case, see~\cite{FS}.}
we find
\eq
G_{(e^{i\lambda})}[a] =
\left({e_\star}^{\delta_\lambda} \star {e_\star}^{-\delta_\lambda + i \Lambda_\lambda[a]}
\right) =: {e_\star}^{\Lambda'_\lambda[a]},
\en
where we have introduced the star exponential $e_\star$ by
$({e_\star}^{f}) \star g \equiv (e^{f \star}) g$. \,
$G_{(e^{i\lambda})}[a]$ is clearly $\star$-invertible; its inverse is given
by $G_{(e^{-i\lambda})}[a + d\lambda] = {e_\star}^{-\Lambda'_\lambda[a]}$.

The $G_g[a]$ play the role of ``noncommutative group elements''.\footnote{It
would be nice to get a better understanding of
the relation to~\cite{Harvey},
where the gauge group of noncommutative gauge theory was
studied in the operator formalism.}
As in the case of $\star$ and $\DD_{[a]}$ the existence of $G_g[a]$ is
more important for the following then the explicit formula for it.

\section{Noncommutative line bundle}

\subsection{Classical line bundle}

Let us recall that a  classical (complex) line bundle is uniquely determined by
a covering $\{ U_k\}$
of a Manifold $M$ and a collection
of transition functions 
$g_{jk} \in C_{\BC}^\infty(U_j \cap U_k)$ 
satisfying relations
\eq
g_{ij} g_{jk} = g_{ik},\label{ggg}
\en
\eq 
g_{jk} g_{kj} = 1 ,
\en
on all
intersections $U_i \cap U_j \cap U_k$ and
$U_j \cap U_k$ respectively. (Here and in the
following there is no sum
over repeated indices.) 

A collection of
functions $\psi_k \in C_{\BC}^\infty(U_k)$ 
satisfying
\eq
\psi_j = g_{jk} \psi_k
\en
on the overlaps $U_j \cap U_k$ define a section $\psi = (\psi_k)$.
A set of local 1-forms $a_k$, satisfying
\eq
a_j = a_k + \ed \lambda_{jk}, \qquad
\ed \lambda_{jk} \equiv i g_{jk} \ed g_{kj} \label{connection}
\en
defines a connection on the line bundle.
{}From (\ref{ggg}) it follows that
\eq
\lambda_{ij} + \lambda_{jk} +\lambda_{ki} = 2 \pi n,  \label{chern}
\quad n \in \BZ .
\en
A connection always exists by the following construction: 
Consider a partition of unity $\{h_k^2\}$,
\eq
\sum_k h_k^2 = 1,
\en
subordinate to
the covering $\{U_k\}$ and define local 1-forms
\eq
a'_j = \sum_k i h_k^2 g_{jk} \ed g_{kj}. \label{localform}
\en
These satisfy (\ref{connection}) by virtue of (\ref{ggg}). 
The cohomological class $[\ed a'] = [\ed a]$ in $H^2(M)$ is the
Chern class of the line bundle and is characterized by the
number $n$ in (\ref{chern}).

Two line bundles $\{\tilde g_{jk}\}$, $\{g_{jk}\}$ 
are equivalent if there
exists a collection of non-vanishing functions 
$f_k \in C_{\BC}^\infty(U_k)$
such that
\eq
\tilde g_{jk} = \zeta_j g_{jk} \zeta^{-1}_k.
\en
The corresponding expressions for the sections and gauge potentials
are
\eq
\tilde \psi_k = \zeta_k \psi_k, \qquad \tilde a_k = a_k + i \zeta_k \ed \zeta_k^{-1}.
\en

\subsection{Noncommutative transition functions}

The finite noncommutative gauge transformations
introduced in section~\ref{sec2} depend on a
local gauge potential defined in an open subset
of $\BR^n$. Assuming that we have the structure of a
classical line bundle on a given Poisson manifold $M$
we can use the consistency condition to construct a
``noncommutative'' line bundle -- the quantization
(in the sense of deformation quantization) of the
given line bundle.
We choose a  covering $\{U_k\}$ of $M$ such
that the patches and all their non-empty intersections
are diffeomorphic to $\BR^n$.
The $C^\infty$-functions on all these open
subsets of $M$ 
become formal power series in a deformation parameter.
Obviously it makes sense to consider the restriction of
the star product to any patch or any intersection
of patches.

Choosing $g_1 = g_{ij}$, $g_2 = g_{jk}$, $g_1 \cdot g_2
= g_{ik}$ and $a = a_k$ in the consistency relation (\ref{consistency})
gives the following relation in the intersection
$U_i \cap U_j \cap U_k$: 
\eq
G_{g_{ij}}[(a_k)_{g_{jk}}]\star G_{g_{jk}}[a_k] =
G_{g_{ij}}[a_j] \star G_{g_{jk}}[a_k] = G_{g_{ik}}[a_k]
\en
where we have used $(a_k)_{g_{jk}} = a_k + d \lambda_{jk} = a_j$.
For the special case $g_1 = g_{kj}$, $g_2 = g_{jk}$ with
$g_{kj} g_{jk} = 1$ we find an expression for the inverse
of $G_{g_{jk}}[a_k]$:
\eq
G_{g_{kj}}[(a_k)_{g_{kj}}]\star G_{g_{jk}}[a_k] =
G_{g_{kj}}[a_j] \star G_{g_{jk}}[a_k] = 1
\en
Similarly in the gauge equivalence relation (\ref{equivalence})
put $g = g_{jk}$ and $a = a_k$,
then
\eq
\DD_{[(a_k)_{g_{jk}}]} \star G_{g_{jk}}[a_k] =
\DD_{[a_j]} \star G_{g_{jk}}[a_k] = G_{g_{jk}}[a_k]\star
\DD_{[a_k]}.
\en
It is consistent to use an abbreviated notation
\eq
G_{jk} \equiv G_{g_{jk}}[a_k], \qquad
\DD_k \equiv\DD_{[a_k]}.
\en 
The fundamental relations
are then
\eq
G_{ij} \star G_{jk} = G_{ik} , \qquad
G_{kj} \star G_{jk} = 1 \label{nclinebundle}
\en
and
\eq
\DD_j \star G_{jk} = G_{jk} \star \DD_k .
\en
(There is no summation over $j$ or $k$ in these formulas.)

In view of (\ref{nclinebundle}), the $G_{jk}$
play the role of noncommutative transition functions.
The collection of $G_{jk} \in C_{\BC}^\infty(U_j \cap U_k)[[\hbar]]$
is a good candidate for a
noncommutative line bundle in the sense
of deformation quantization. As we will see later the $D_k$ play the
role of a local connection.

We shall now study the freedom in the construction of the $G_{jk}$.
The $G_{jk}$ depend explicitly on a classical connection $a_k$.
For a given classical line bundle, i.e.\ fixed $g_{jk}$,
the  choice of different $a_k$ only changes the star product
on to one in the same equivalence class. The reason is that
the new $a_k$ differ from the old ones by a global one-form $b$.
The equivalence is given by $\DD[b]$.
For an equivalent classical line bundle with
transition functions 
$\widetilde g_{jk} = \zeta_j g_{jk} \zeta^{-1}_k$
and local connection
forms $\widetilde a_k = a_k + d \zeta_k$ we find new transition functions for the
noncommutative line bundle of the form
\eq
\widetilde G_{jk} = G_{\zeta_j}[a_j] \star G_{jk} \star (G_{\zeta_k}[a_k])^{-1}.
\en
Here we have twice used the consistency relation (\ref{consistency}).
The additional freedom from deformation quantization is that we should
consider equivalence classes of star products on each patch.
This is related to the further
requirement that the equivalence classes of noncommutative
line bundles should be independent
of the choice of coordinates in the individual patches~\cite{JSW2}.
Such coordinate changes in patches $U_j$ and $U_k$ induce equivalence
maps $\Sigma_j$, $\Sigma'_k$ such that
$\mathrm{Ad}_\star G_{jk}$ becomes $\Sigma_j\circ\mathrm{Ad}_\star
G_{jk}\circ(\Sigma'_k)^{-1}$.
The classical freedom discussed above is included as a special case.

\subsection{Quantized line bundle}

We have explicitly constructed a candidate for a noncommutative
line bundle starting from a classical line bundle and a solution
$G_g[a]$ to the consistency relation. 
Now we would now like to collect the essential properties
of a noncommutative line bundle in the
framework of deformation quantization:

Let $\{g_{jk}\}$ be a classical line bundle on a Poisson manifold
$(M,\theta)$ and let~$\star$ be a star product (deformation quantization
of $\theta$.)
We define a noncommutative line bundle $(G,\star)$ to be a set of noncommutative
transition functions $G_{jk}$ of the following form:
\begin{quote}
{\it The $G_{jk} \equiv G_{g_{jk}}[a_k] \in C^\infty_\BC(U_j \cap U_k)[[\hbar]]$
are local functions
of the classical transition functions $g_{jk}$ and of
a given connection $a_k$ and their derivatives satisfying
\eq
G_{ij} \star G_{jk} = G_{ik}, \qquad G_{kk} = 1, \label{transition}
\en
with the commutative limit $(\theta=0)$: $G_{jk} \rightarrow g_{jk}$.}
\end{quote}
\begin{quote}
{\it
We call two line bundles $(G,\star)$, $(G',\star')$ equivalent if
they are based on equivalent classical line bundles, $g_{jk} \sim g'_{jk}$,
and there
are equivalence maps $\Sigma_k$ on every patch $U_k$ relating $\star$ and
$\star'$.}
\end{quote}
(Here $\star$ and $\star'$ can  possibly be quantizations
of two different Poisson structures $\theta$ and $\theta'$.)

In place of connection forms we introduce
invertible maps $\DD_k$ that define covariant functions
$\DD_k(f)$ on every patch:
\begin{quote}
{\it The $\DD_k \equiv \DD_{[a_k]}: C^\infty_\BC
(U_k)[[\hbar]]\rightarrow
C^\infty_\BC(U_k)[[\hbar]]$ are  local functions of the given connection
$a_k$ and its derivatives, with the properties that
\eq
\DD_j(f) \star G_{jk} = G_{jk} \star \DD_k(f),\label{covariance}
\en
(or, equivalently, $\mathrm{Ad}_\star G_{jk} = \DD_j \circ \DD_k^{-1}$)
and that
\eq
\DD_k(f\star' g) = \DD_k f \star \DD_k g \label{starprime}
\en
defines a new star product $\star'$. Let us note that if this is the
case,
$\star'$ is independent of the patch $U_k$, i.e., it depends on the $a_k$ only
via the global 2-form $f = d a$.
}
\end{quote}

Choosing appropriate representatives in the equivalence classes
one can introduce the tensor product
of two noncommutative line bundles. Namely
\eq
G''_{jk} = \DD_k(G'_{jk}) \star G_{jk} \label{tensor}
\en
satisfies again the consistency condition
with star product $\star$. Here $G'_{jk} = G'_{g'_{jk}}[a'_k]$ satisfies
the consistency condition with the star product $\star'$ (\ref{starprime}).
The line bundle based on $G''$ is equivalent to the one based on $G_{g' g}[a' +
a]$.

\section{Sections}

A section $\Psi =(\Psi_k)$ is a collection of functions 
$\Psi_k \in C_{\BC}^\infty(U_k)[[\hbar]]$ satisfying consistency
relations
\eq
\Psi_j = G_{jk} \star \Psi_k \label{section}
\en
on all intersections
$U_j \cap U_k$. With this definition the space of sections
$\EE$ is a right $\Ax$ module. We shall use the notation
${\cal E}_{\Ax}$ for it. The right action of the function $f\in \Ax$
is the regular one
\eq
\Psi. f = (\Psi_k \star f).
\en
Using the maps $\DD_k$ it is easy to turn ${\cal E}$ into a left
${\Ax}' =(C^\infty_\BC(M)[[\hbar]],\star')$ 
module ${}_{{\Ax}'}{\cal E}$. The left action of ${\Ax}'$ is given by
\eq  
f .\Psi = (\DD_k(f) \star \Psi_k). \label{leftaction}
\en
It is easy to check using  (\ref{covariance}) that the left action is
compatible with (\ref{section}), in fact this was the reason why
we introduced covariant functions in the first place.
From  the property (\ref{starprime})
of the maps $\DD_k$ we find 
\eq
f.(g.\Psi)) = (f \star' g).\Psi .
\en 
Together we have a bimodule structure ${}_{{\Ax}'}{\cal E}_{\Ax}$
on the space of sections. 

\section{Connection}

Let us note that using the Hochschild complex we can introduce a natural differential
calculus on the algebra $\Ax$. The $p$-cochains, elements of $C^p=\mbox{Hom}_{\BC}({\Ax}^{\otimes p},
\Ax)$
play the role of $p$-forms and the derivation $\ed: C^p \rightarrow C^{p+1}$ is given on 
$C\in C^p$ as
\eqa
\lefteqn{(\ed C)(f_1,f_2,\ldots,f_{p+1})  =  f_1 \star C(f_2,\ldots,f_{p+1}) - 
C(f_1\star f_2,\dots,f_{p+1})}\nn
&& {} + C(f_1,f_2\star
f_3,\ldots,f_{p+1})-\ldots+(-1)^p C(f_1,f_2,\ldots,f_p\star f_{p+1})\nn
&& {} +(-1)^{p+1} C(f_1,f_2,\ldots,f_p)\star f_{p+1}. \label{der}
\ena
We also have the cup product $C_1 \cup C_2 $ of two cochains 
$C_1\in C^{p}$ and $C_2\in C^{q}$;
\eq
(C_1\cup C_2)(f_1,...,f_{p+q})=C_1(f_1,...,f_p)\star C_2(f_{p+1},...,f_q).
\en
The cup product extends to a map from $({\cal E}
\otimes_{\Ax} C^p) \otimes_{\Ax} C^q$ to ${\cal E} \otimes_{\Ax} C^{p+q}.$ 

A connection $\nabla : {\cal E}\otimes_{\Ax} C^p 
\rightarrow {\cal E}\otimes_{\Ax} C^{p+1}$ can now be defined by a formula
similar to (\ref{der}) using the natural extension of the
left and right module structure of $\EE$ to ${\cal E}\otimes_{\Ax}  C^p$. 
Namely, for a $\Phi\in {\cal E}\otimes_{\Ax} C^p$ we have
\eqa
\lefteqn{(\nabla \Phi)(f_1,f_2,\ldots,f_{p+1}) = f_1.\Phi(f_2,\ldots,f_{p+1}) - 
\Phi(f_1\star f_2,\ldots,f_{p+1})} \nn
&& {} + \Phi(f_1,f_2\star
f_3,\ldots,f_{p+1})-\ldots+(-1)^p\Phi(f_1,f_2,\ldots,f_p\star f_{p+1}) \nn
&& {} +(-1)^{p+1} \Phi(f_1,f_2,\ldots,f_p).f_{p+1}. \label{c}
\ena
It is easy to check that $\nabla$ satisfies the graded Leibniz rule
with respect to the cup product and thus 
defines a bona fide connection on the module ${\cal E}_{\Ax}$.
On the sections the connection $\nabla$ introduced here is simply the
difference between the
two actions of $C^{\infty}_{\BC}(M)[[\hbar]]$ on $\EE$:
\eq
(\nabla \Psi)(f) = f.\Psi - \Psi.f = (\DD_k(f)\star\Psi_k - \Psi_k \star f).
\en
In terms of $\AA_k \equiv \DD_k - \id$ this reads
$\nabla\Psi = (d \Psi_k + \AA_k \star \Psi_k)$.

A simple computation using the definition (\ref{c}) reveals that the corresponding
curvature $K_\nabla \equiv \nabla^2 :{\cal E}\otimes_{\Ax} C^p \rightarrow {\cal
E}\otimes_{\Ax} C^{p+2}$ measures the difference between the two star products $\star'$ and
$\star$. On a section $\Psi$ it is given by
\eq
(K_\nabla \Psi)(f,g)= (\DD_k(f\star'g- f\star g)\star\Psi_k) \label{curv} 
\en

\section{Projective module}

Now we show that in the case of a
compact Poisson manifold $M$
the space of sections $\EE$  as a right $\Ax$-module
is projective of finite type. We need to introduce a
\emph{$\star$-covariant partition of unity}:
We start with a partition of unity $\sum_i\rho_i =1$,
$\rho_i \in C^\infty(M)$, subordinate to the
finite covering $\{U_i\}_{i=1}^n$ of $M$.
We define functions
$h_i \in C^\infty_\BC(M)[[\hbar]]$ with support in $U_i$
as $h_i = \sqrt[\star']{\rho_i}$.
In the sense of formal
power series the $h_i$
are \mbox{$\star'$-in}\-ver\-tible on $U_i$.
On any patch $U_k$ 
we have the following property 
\eq
\sum_i \DD_k(h_i) \star \DD_k(h_i) =1. \label{pou}
\en
There is actually no need to know the maps $\DD_k$ on all of $\Ax$. It is
enough to know the family of functions $H_{kj} = \DD_k(h_j)$ that feature
in the $\star$-covariant partition of unity. These functions can also be
introduced abstractly by the following two requirements
for all $k$
\eq
\sum_j H_{kj} \star H_{kj} = 1 ,
\qquad
H_{ik} \star G_{ij} = G_{ij} \star H_{jk}. \label{HG}
\en
Maps $\widetilde\DD_k$ with property (\ref{covariance}) can always be
found by a formula analogously to (\ref{localform}):
\eq
\widetilde\DD_k(f) = \sum_j H_{kj}\star H_{kj} \star G_{kj} \star f \star
G_{jk}.
\en
Equation (\ref{HG}) together with (\ref{transition}) can be taken as a starting point
for an abstract definition of a noncommutative line bundle even for
an associative algebra which not necessarily given by a star product.

Next we define a right $\Ax$-module morphism $\epsilon: \EE \rightarrow \Ax^n$.
It sends  a section $\Psi = (\Psi_j)$ to the $n$-tuple $(t_j)$ with
\eq
t_j = \DD_j(h_j) \star \Psi_j \quad \in \Ax. \label{eps}
\en
There is also a surjection $\pi: \Ax^n \rightarrow \EE$
which sends the $n$-tuple $(t_k)$ to the section $\Psi = (\Psi_k)$
with
\eq
\Psi_k = \sum_j G_{kj} \star \DD_j(h_j)  \star t_j.
\en
This is obviously also a right $\Ax$-module morphism and it is
also easy to  check that
$\Psi_k$ thus defined satisfies (\ref{section}).
The composition $\pi\circ\epsilon$ is the identity on $\EE$
as we can check using (\ref{pou}):
\eqa
\lefteqn{\sum_j G_{kj} \star \DD_j(h_j)  \star \DD_j(h_j) \star \Psi_j
 = \sum_j  \DD_k(h_j)  \star \DD_k(h_j) \star G_{kj} \star \Psi_j}\nn
&& = \sum_j  \DD_k(h_j)  \star \DD_k(h_j) \star \Psi_k
= \Psi_k. \phantom{\DD_k(h_j)  \star \DD_k(h_j) \star \Psi_k}
\nonumber
\ena
The opposite composition $\epsilon\circ\pi: \Ax^n\rightarrow \Ax^n$ is the projection
defined by the projector $P$ 
\eq
P_{ij} = \DD_i(h_i)\star G_{ij}\star \DD_j(h_j), \label{projector} 
\en
with
\eq
P_{ij}=\sum_k P_{ik}\star P_{kj}.
\en
This makes $\EE_\Ax$ to a projective right $\Ax$-module of finite type.
The sections $\Psi \in \EE$ can now be identified through the embedding (\ref{eps}) 
with elements $t=(t_j)$
of the free module $\Ax^n$ satisfying
\eq
\sum_j P_{ij}\star t_j =t_i.
\en
Within this identification the right $\Ax$ and the left $\Ax'$ actions on $\epsilon(\EE)\subset \Ax^n$
look like
\eq
t.f=(t_j \star f) \label{rigtont}
\en
and
\eq
f.t=((\DD_j(h_j)\star \DD_j(f) \star \DD_j(h_j)^{-1})\star t_j).\label{leftont}
\en

Let us note that $\EE$ naturally has also the structure of a left projective $\Ax'$-module.
Namely all the construction above can be easily modified as follows. We start with an equivalent
line bundle $(G'_{jk}=\DD_j^{-1}(G_{jk}), \star')$ and take for sections of this line bundle
collections of functions 
\eq
\Psi'=(\DD_k^{-1}(\Psi_k)).
\en 
Thus 
\eq
\Psi'_k= \Psi'_j \star' G'_{jk}.
\en
This gives an equivalent description of
${}_{\Ax'}\EE_{\Ax}$ as a bimodule. 
Using now a  $\star'$-covariant partition of unity
$h'_k$  we have the projector 
\eq
P'_{jk}=\DD_j^{-1}(h'_j)\star'G'_{jk}\star'\DD_k^{-1}(h'_k)
\en 
and the identification ${}_{\Ax'}\EE \cong {\Ax'}^{n}P'$ of left $\Ax'$ modules.

\section{Morita equivalence}

According to the definition of the two star products
$\star$ and $\star'$, these are equivalent in each individual patch $U_k$.
Obviously they are equivalent on the whole of $M$ in the special case if the
classical line bundle $\{g_{ij}\}$ that we take as a starting point
 is trivial. Then there is a globally defined connection 
one-form
$a$ and consequently a globally defined covariantizing map $\DD$. 
We recall that in the case of a line bundle which is based on the particular choice
of transition functions $G_{ij}$ described in section~\ref{sec2}, $\star'$
is the deformation quantization
of the Poisson tensor
$\theta' = \theta - \theta f\theta + \theta f\theta f \theta \mp \ldots$.

Here we want to show that in the case of a general line bundle $\{g_{ij}\}$ the two star products 
$\star$ and $\star'$ define Morita equivalent algebras.
For this let us introduce a new space of sections $\overline{\EE}$.
Sections in $\overline{\EE}$ are collections of functions
$\overline{\Psi} =(\overline{\Psi}_k)$ 
satisfying the opposite consistency relations
\eq
\overline{\Psi}_j = \overline{\Psi}_k \star G_{kj}\label{overlinesection}
\en
on all intersections $U_j \cap U_k$. 
There are left module morphisms $\overline{\epsilon}: \overline{\EE} \rightarrow \Ax^n$
and $\overline{\pi}: \Ax^n \rightarrow \overline{\EE}$
given by
\eq
(\overline{\psi}_j) \stackrel{\overline{\epsilon}}{\mapsto} (\overline{t}_j) =
(\overline{\psi}_j \star\DD_j(h_j)),
\en
\eq
(\overline{t}_k) \stackrel{\overline{\pi}}{\mapsto}
(\overline{\psi}_k) = (\sum_j\overline{t}_j \star\DD_j(h_j)\star G_{jk}).
\en
The sections in $\overline{\EE}$ can be identified 
with elements $\overline{t}=(\overline{t}_j)$
of the free module $\Ax^n$ satisfying
$\sum_i \overline{t}_i \star P_{ij} =\overline{t}_j$.
Clearly with this definition the roles of 
the algebras $\Ax$ and $\Ax'$
interchange and the space of sections
$\overline{\EE}$ becomes a ${}_{\Ax}\overline{\EE}_{\Ax'}$ bimodule.
One way to show the Morita equivalence of $\Ax$ and $\Ax'$ is to prove that
\eq
{}_{\Ax'} \EE \otimes_{\Ax} \overline{\EE}_{\Ax'} \cong {}_{\Ax'} {\Ax'}_{\Ax'}
\label{Morita1} 
\en
and 
\eq
{}_\Ax \overline{\EE} \otimes_{\Ax'} \EE_\Ax \cong {}_\Ax \Ax_{\Ax} \label{Morita2}
\en
as bimodules. 
This will be done in the rest of this section.

We start with the following observation: For any $M \in \mathrm{M}_n(\Ax)$,
\eq
\sum_{ij} P_{ri} \star M_{ij} \star P_{js}
= \DD_r(h_r) \star \DD_r(f_M) \star \DD_r(h_r)^{-1} \star P_{rs},
\label{oben1}
\en
with the globally defined function
\eq
f_M = \sum_{ij} \DD_i^{-1}\Big(\DD_i(h_i)\star M_{ij} \star \DD_j(h_j) \star
G_{ji}\Big) \quad \in \,C^\infty_\BC(M)[[\hbar]].
\en
The proof uses $G_{js} = G_{ji}\star G_{ir} \star G_{rs}$ in $U_j\cap U_i\cap
U_r\cap U_s$ and
$G_{ri} \star (\ldots)\star G_{ir} = \DD_r(\DD_i^{-1}(\ldots))$.
Similarly
\eq
\sum_{ij} P_{ri} \star M_{ij} \star P_{js}
= P_{rs}\star\DD_s(h_s)^{-1}\star\DD_s(f_M)\star\DD_s(h_s). \label{oben2}
\en
Note that $P_{rs}$ is a section in $\EE$
for every fixed $s$ under the embedding $\epsilon$.
In this language, (\ref{oben1}) is just the
left $\Ax'$-action (\ref{leftont}) of $f_M$ on this section. Using an analogous formula
for the right $\Ax'$-action on $\overline{\EE}$ we see that (\ref{oben2})
is the right $\Ax'$-action of $f_M$ on $P_{rs}$ viewed as a section
in $\overline{\EE}$ for fixed $r$ under the embedding $\overline{\epsilon}$.
With slight abuse of notation
we can summarize (\ref{oben1}) and (\ref{oben2}) in matrix form as
\eq
f_M . P = P\star M \star P = P. f_M \label{oben3}.
\en
Again under the embeddings $\epsilon$ and $\overline{\epsilon}$,
the tensor product of two arbitrary sections in $\EE$ and $\overline{\EE}$,
respectively, takes the form
\eq
\sum_i P_{ri} \star t_i \otimes_\Ax \sum_j \overline{t}_j \star P_{jm}
= \sum_{ijs} P_{ri} \star t_i\star \overline{t}_j \star P_{js} \otimes_\Ax
P_{sm}.
\en
This implies  that sections in $\EE \otimes_\Ax \overline{\EE}$ have the
form
\eq
\sum_{ijs} P_{ri} \star M_{ij} \star P_{js} \otimes_\Ax P_{sm}.
\en
with $M \in \mathrm{M}_n(\Ax)$ and that
$\Psi^{ij} = (P_{ri}  \otimes_\Ax P_{js})$ is a generating
family for $\EE \otimes_\Ax \overline{\EE}$.
Using (\ref{oben1}) and (\ref{oben2}) in the
form (\ref{oben3}), we find the following chain of equalities  (in matrix notation)
\eq
f_M . P \otimes_\Ax P = P\star M \star P \otimes_\Ax P = P \otimes_\Ax P.f_M
\en
which proofs (\ref{Morita1}) and, in view of the remarks at the end of the previous
section, also (\ref{Morita2}). To summarize, we have shown that the algebras $(\Ax,
\star)$ and $(\Ax',\star')$ are Morita equivalent  by an
explicit construction
of equivalence bimodules $\EE$ and $\overline{\EE}$.

\section{Discussion}

Here we would like to make some comments, without going into details, relating our
discussion of Morita equivalence to earlier
works~\cite{Waldmann}, \cite{Bursztyn} on the deformation of projective modules 
and the Morita equivalence of $\star$-products which are deformation quantizations
of some given Poisson structure $\theta$. For this purpose we have to assume
that the connection $a$ enters all expressions ($G_g[a]$, $\DD_{[a]}$, $\theta'$, \ldots)
in the main body of our paper in the form $\hbar a$.
Furthermore we want to focus on the case where
all constructions of our paper are done with the explicit choice
of noncommutative transition functions $G_{jk}$ and of covariantizing maps
$\DD_k$ as described in \ref{sec2}.
Different choices would lead to projective modules which are 
equivalent in the sense of \cite{Waldmann}. The choice of a different
covariant partition of 
unity would have the same effect. 

In the case of a nontrivial
classical line bundle, the Poisson structures~$\theta$ and
$\theta' = \sum_{n=0}^\infty (-\hbar)^n \theta (f \theta)^n$
live in  two different equivalence classes in the sense of
Kontsevich~\cite{Kontsevich}.\footnote{For simplicity we have written the formula
for $\theta'$ for some choice of local coordinates, matrix multiplication is implied
 and $f$ is the coefficient matrix of the curvature form $\ed a$.}
 Let us recall that equivalence classes of star products are in 
one to one correspondence with equivalence classes of Poisson structures.
In the case of a trivial line bundle $\theta$ and
$\theta'$ are equivalent.

We can now modify the action \cite{Bursztyn} of the Picard group
$\mbox{Pic}(C^\infty(M))\cong
\mbox{Pic}(M) \cong H^2(M,\BZ)$ on the equivalence classes of
the star products quantizing $\theta$.
An element $f\in H^2(M,\BZ)$ simply sends the equivalence class $[\star]$
of star products corresponding to the
equivalence class of Poisson structures $[\theta]$ to the
equivalence class $[\star']$ corresponding to the
equivalence class of Poisson structures
$[\theta']$.
With the modified action of $\mbox{Pic}(M)$ two star-products quantizing the same Poisson 
structure are also Morita equivalent if and only if
they are related by the action of an element of $\mbox{Pic}(C^\infty(M)).$
This can be shown either directly or follows from the comparison of the action of 
$\mbox{Pic}(C^\infty(M))$ introduced here with the one of the paper \cite{Bursztyn}.
Since our projector $P$, see equation (\ref{projector}),
is a deformation of the classical full projector
$P_0 = P(\hbar=0)$ it is also a full one \cite{Waldmann}. We can use it in the construction of the 
$\mbox{Pic}(C^\infty(M))$ 
action of \cite{Bursztyn}. When we do this the two actions when applied to $\star$ give equivalent star
products on $M$. The corresponding equivalence map $\widetilde{\;}$ sends the function 
$f$ to the function $\tilde f = \sum_i \DD_i^{-1}(\DD_i(h_i)\star f \star \DD_i(h_i))$.
The function $\tilde f$ is simply the function
$f_M$ of the previous section for the diagonal 
matrix
$M=\mbox{diag}(f,\ldots,f)$.

Let us finish with a note concerning an
alternative proof of Morita equivalence of $\Ax$ and $\Ax'$.
From the discussion of the previous section it also
follows that $\Ax' \cong P\star M(A)\star P$ with the
full projector $P$. This is just another way to express
the Morita equivalence of $\Ax$ and $\Ax'$.

\subsection*{Acknowledgement}

B.J.\ would like to thank A.\ Isaev for an inspiring question.

\end{document}